*Research Article*

# Observing and Reducing IFUs: INTEGRAL and PMAS—Properties of the Ionized Gas in HH 202

**Luis López-Martín**[1,2]

[1] *Instituto de Astrofísica de Canarias, La Laguna, 38205 Tenerife, Spain*
[2] *Departamento de Astrofísica, Universidad de La Laguna, La Laguna, 38206 Tenerife, Spain*

Correspondence should be addressed to Luis López-Martín; luislm@iac.es

The reduction of integral field spectroscopy (IFS) data requires several stages and many repetitive operations to convert raw data into, typically, a large number of spectra. Instead there are several semiautomatic data reduction tools and here we present this data reduction process using some of the Image Reduction and Analysis Facility (IRAF) tasks devoted to reduce spectroscopic data. After explaining the whole process, we illustrate the power of this instrumental technique with some results obtained for the object HH202 in the Orion Nebula (Mesa-Delgado et al., 2009).

## 1. Introduction

Simultaneously storing both spectral and spatial information, 3D spectroscopy (known also as 2D spectroscopy, spectral imaging, or integral field spectroscopy) guarantees the homogeneity of the data, offers a perfect way to address astrophysical problems, and opens up new lines of research. Since its inception in the eighties (Vanderriest [1] on the CFHT) and early nineties, research in this field has grown enormously. Three-dimensional spectrographs, or integral field units, (IFUs) provide spectra for a large number of spatial elements ("spaxel") within a two-dimensional field of view, rather than only along a traditional one-dimensional spectrograph slit. Figure 1 shows how this spectroscopical information is collected into a data cube. Most of the advantages of the IFS technique are direct consequences of the simultaneity when recording spatial and spectral information, which guarantees a great homogeneity in the data. The complexity of these data makes it a little more difficult to reduce them than in long slit spectroscopy; in next sections we will briefly describe how it is possible do this reduction using IRAK tasks. To finish, we will show a scientific application of this observational technique: 3D spectroscopy of photoionized HH objects in the Orion nebula—the case of HH 202—[2]. Part of this work was contributed in the meeting "Metals in 3D: New insights from Integral Field Spectroscopy," held in Granada in 18–20/04/2012.

## 2. Observing with INTEGRAL (WHT) and PMAS (CAHA)

We will describe the process of reduction of 3D spectroscopic data through spectra obtained with two different instruments located at WHT 4.2 m (Observatorio del Roque de los Muchachos) and CAHA 3.5 m (Observatorio de Calar Alto) telescopes: INTEGRAL and PMAS.

INTEGRAL [3], is mounted on the Nasmyth focus of the WHT at the Roque de los Muchachos Observatory on La Palma. In its standard configuration, it has three fiber bundles that simultaneously feed the entrance of the WYFFOS fibre spectrograph. The three bundles are located in the focal plane on a revolving wheel (see Figure 2).

But currently four bundles are mounted in the INTEGRAL Swinging Plate: SB1, SB2, SB3, and equalized. At the focal plane the fibers are arranged in two groups, one forming a rectangle, while the other forms a ring (except for the equalized bundle) which is intended for collecting background light (for small-sized objects). Table 1 shows the main characteristics of the four bundles.

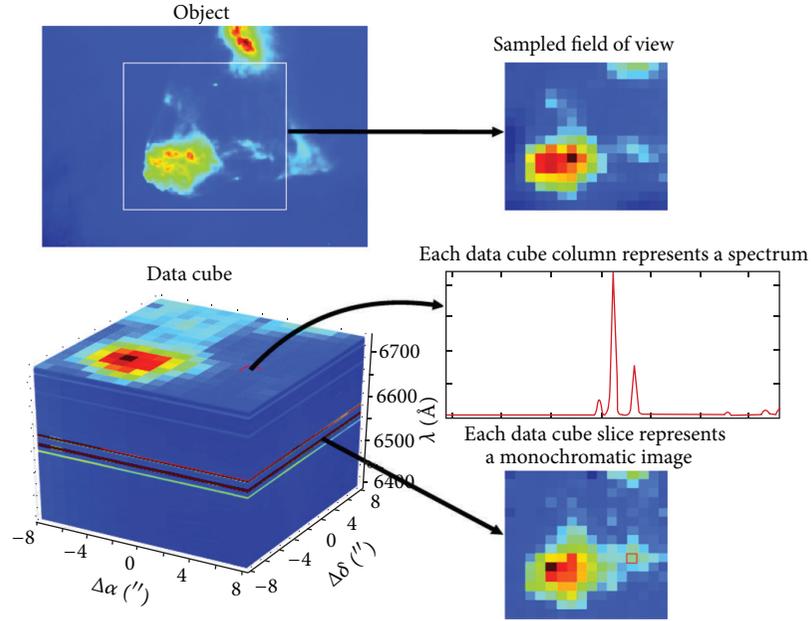

FIGURE 1: Section of a Hα *HST* image of Orion [4] with a sampling of 1 arcsec (top left, white box indicates the FoV of a typical 2D PMAS observation). A result of this PMAS observation with spaxels of 1 × 1 arcsec is shown on the top right. Bottom left shows a section of the data cube observed where we can see a monochromatic image at the upper slice. Bottom right shows an integrated spectrum of a group of spaxels and a flux map integrated over an emission line (figure extracted from the Ph.D. thesis of M. Núñez-Díaz).

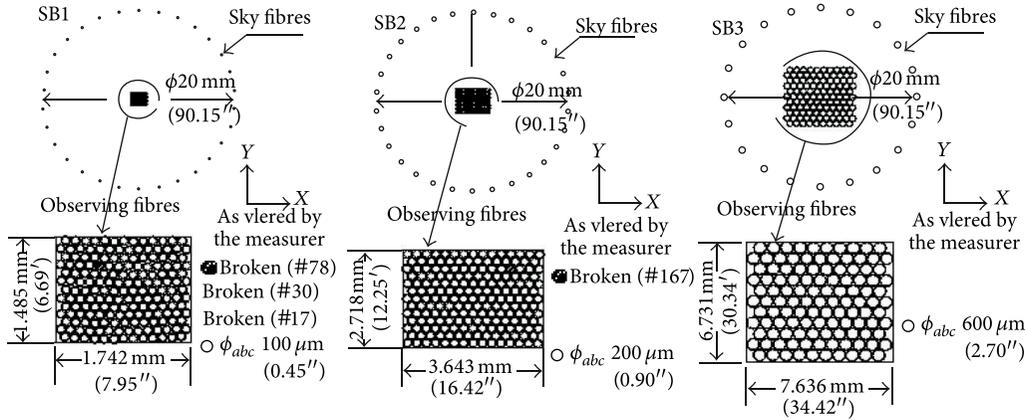

FIGURE 2: Sketch of the geometrical characteristics of the bundles of fibres in INTEGRAL. Extracted from Arribas et al. [3].

Their tilt angles can be varied in order to select a specific wavelength region. For any particular grating the spectral resolution depends on the fibre bundle as a consequence of the different fibre sizes. Table 2 shows the spectral resolution achieved with the INTEGRAL bundles for the available gratings.

PMAS [5], the Potsdam Multi-Aperture Spectrophotometer, is an integral field instrument developed at the AIP. It is currently installed at the Calar Alto Observatory 3.5 m Telescope in Spain. Actually, it is an integral field spectroscopy with fiber-coupled lens array or fiber bundle IFU and fiber spectrograph.

PMAS employs an all-refractive fiber spectrograph, built with $CaF_2$ optics, to provide good transmission and high image quality over the entire nominal wavelength range. A set of user-selectable reflective gratings provides low to medium spectral resolution in first order of approximately 1.5, 3.2, and 7 Å, depending on the groove density (1200, 600, 300 gr/mm).

The instrument was specifically designed to address the science case of 3D spectrophotometry of spatially resolved, individual objects, with an emphasis on broad wavelength coverage in the optical wavelength regime. Table 3 lists the main properties of the PMAS spectrograph. In addition, the properties of the gratings available for PMAS are listed in Table 4, together with the achievable spectral resolutions.

Table 1: Characteristics of the fiber bundles of INTEGRAL (see http://www.iac.es/proyecto/integral/).

| | Fiber size ($''$) | Number (field + sky) | Spatial covering ($''$) | Fiber size (pixeles) | External ring ($''$) |
|---|---|---|---|---|---|
| SB1 | 0.45 | 205 (175 + 30) | 7.8 × 6.4 | 1.9–2.2 | 90 |
| SB2 | 0.90 | 219 (189 + 30) | 16.0 × 12.3 | 3.5–4 | 90 |
| SB3 | 2.70 | 135 (115 + 20) | 33.6 × 29.4 | 13–14 | 90 |
| "Equalized" | 0.45 | 115 (115 + 0) | 6.3 × 5.4 | 1.9–2.2 | NO |

Table 2: Mean spectral resolution, linear dispersions, and spectral coverage for different gratings and bundles (see http://www.iac.es/proyecto/integral/).

| | 1200 L/mm | 600 L/mm | 300/316 L/mm |
|---|---|---|---|
| Resolution (SB1) (Å) | 0.7 | 1.3 | 2.6 |
| Resolution (SB2) (Å) | 1.3 | 2.6 | 5.2 |
| Resolution (SB3) (Å) | 5.5 | 11 | 22 |
| Dispersion (Å/pix) | 0.4 | 0.8 | 1.6 |
| Covering (Å) | 1620 | 3240 | 6480 |

Table 3: Characteristics of PMAS lens array (see http://www.caha.es/pmas/PMAS_OVERVIEW/pmas_overview.html).

| Standard-IFU | |
|---|---|
| Principle of operation | Square lens array with fore optics |
| Lens array | 16 × 16 square elements, 1 mm pitch |
| | (32 × 32 array upgrade in preparation) |
| 3 magnifications | 0.5 arcsec sampling, 8 × 8 arcsec$^2$ FOV |
| | 0.75 arcsec sampling, 12 × 12 arcsec$^2$ FOV |
| | 1.0 arcsec sampling, 16 × 16 arcsec$^2$ FOV |
| Fiber configuration | 256 OH-doped fibers, 150 um core diameter |

## 3. Reducing IFUs with IRAF

*3.1. Reduction Steps versus IRAF Tasks.* Although there are some 3D reduction packages (R3D [6, 7], P3D [8], etc.), we will describe the data reduction using IRAF reduction package SPECRED. Specific reduction packages probably are faster ways to reach reduced data (once the user knows how it works). Reducing using individual IRAF tasks probably takes a lot of time, mainly because for each reduction stage a different task must be used, and the output of each of these stages must be verified; but on the other hand it can be checked that every task is producing satisfying results.

The data reduction process has several steps that we can summarize in this way: after bias subtraction, spectra are traced on the continuum lamp exposure obtained before each science exposure and wavelength calibrated using a Hg-Ne arc lamp. The continuum lamp and sky flats are used to determine the response of the instrument for each fibre and wavelength. Finally, for the standard stars we coadded the spectra of the central fibres and compared them with the tabulated one-dimensional spectra. Due to the complexity of these data we are going to describe with more detail step by step the process. In Figure 3 we present a scheme of this process in which we can see the products after each step.

*3.2. Bias Subtraction.* The subtraction of the bias level is the first step in the data reduction. This bias level is introduced onto the CCD chip for ensuring that the chip is working in a linear regime.

This pedestal level (bias) can be removed from images according to one of the following proceedings: get the mean value on the overscan region of each frame and subtract this constant off the whole rest of the frame, or average a sample (around 10) of bias frames pixel by pixel in order to get a mean bias frame, and subtract this frame from each image.

In these data we have bias subtracted following the second way using *imcobine* and *imarith* tasks of IRAF.

*3.3. Spectra Extraction.* The extraction of one-dimension spectra from the two-dimensional image is a multistage process. First, we must find the spectra from the image using the continuum lamp as reference. This continuum lamp traces spectra over the CCD image. The task apall just looks over this image and extracts the place where spectra will be found (see Figure 4). This task produces a final multispectra image (with a.ms extension file).

*3.4. Wavelength Calibration.* Now, we would like to see the wavelength calibrated spectra. For this, we need to determine the dispersion solution in order to transform the pixel to the wavelength scale. This can be done interactively for the first time in one calibration lamp reference spectrum (identify task), and this solution is used as a starting point to determine the dispersion solution for the rest of the spectra (reidentify task).

In Figure 5(a) we can see a wavelength calibrated spectrum identifying lamp lines using a database. This identification gives an initial dispersion solution able to be applied to the rest of spectra (Figure 5(b)). In Figure 6 we can see the image resulting to apply the wavelength calibration.

*3.5. Instrumental Response.* Once spectra have been wavelength-calibrated, it is necessary to correct from the instrumental response. This response varies from pixel to pixel not only for a different sensitivity to the intensity of the incoming light but also to the response at a different wavelength.

For both responses we must use continuum lamp and sky spectra, obtained before each object exposure. Using *msresp1d* IRAF task we can obtain that function (see Figure 7).

Table 4: Linear dispersions and spectral coverage for different gratings used in PMAS (see http://www.caha.es/pmas/PMAS_OVERVIEW/pmas_gratings.html).

| Grating design. | Cartr. ID number | Grooves per mm | Dispersion (A/pix) | Blaze angle | Lambda (nm) | dLambda (A) |
| --- | --- | --- | --- | --- | --- | --- |
| U1200 | 1 | 1200 | 0.39 | 10.4 | 300 | 794 |
| V1200 | 2 | 1200 | 0.35 | 17.5 | 500 | 725 |
| R1200 | 3 | 1200 | 0.30 | 26.7 | 750 | 609 |
| I1200 | 4 | 1200 | 0.22 | 36.8 | 1000 | 460 |
| J1200 | 5 | 1200 | 0.22 | 46.0 | 1200 | 341 |
| J1200 | 5 | 1200 | 0.17 | 46.0 | 600 | 450 |
| U600 | 6 | 600 | 0.81 | 5.2 | 300 | 1656 |
| V600 | 7 | 600 | 0.80 | 8.6 | 500 | 1630 |
| R600 | 8 | 600 | 0.75 | 13.9 | 800 | 1533 |
| U300 | 9 | 300 | 1.67 | 2.5 | 300 | 3404 |
| V300 | 10 | 300 | 1.67 | 4.3 | 500 | 3404 |

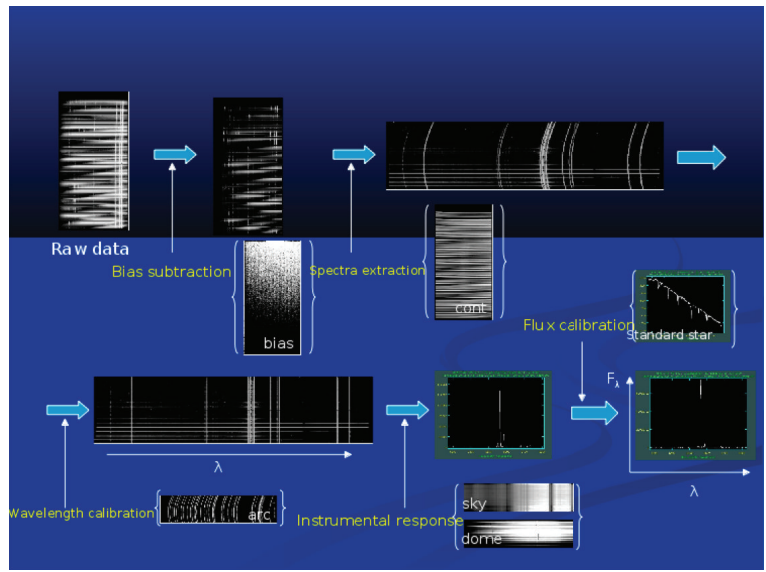

Figure 3: Sketch of the 3D data reduction process. Data taken from INTEGRAL observations.

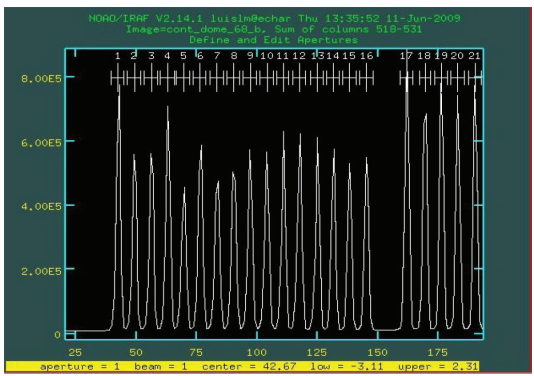

Figure 4: Spectra extraction of INTEGRAL data using continuum lamp.

*3.6. Flux Calibration.* For some users, leaving the wavelength calibrated spectra in terms of integrated numbers of counts is sufficient. For the rest, it is necessary to observe suitable spectrophotometric standard stars, in order to transform the data in flux units.

First, we must find the sensitivity function using the spectrophotometric star spectra with *standard* and *sensfunc* tasks. Then, we need to apply this function to the object spectra (*calibrate* task). This way, we obtain spectra in counts units. At this point, it is necessary to know how much flux units correspond to one count; to do this, we coadd all the spectrophotometric spectra and we compare them with tabulated flux values. In Figure 8(a) we can see how the response function varies with wavelength; once this function is applied we obtain a nonflux calibrated spectrum (Figure 8(b)), and finally if we compare it with tabulated flux values, we obtain the flux calibrated spectra (Figure 8(c)).

## 4. Application of 3D Spectroscopy to HH 202

Perhaps the most interesting feature of 3D spectroscopy is the possibility of having a spectrum for each spaxel in the observed field (typically the fiber in the case of fiber-fed spectrographs as INTEGRAL and PMAS). This means that the spectral features of interest contained in our spectra can be measured throughout the observed field of view and, thus, bidimensional images of these features can be reconstructed by just arranging properly the spectra in the 2D space.

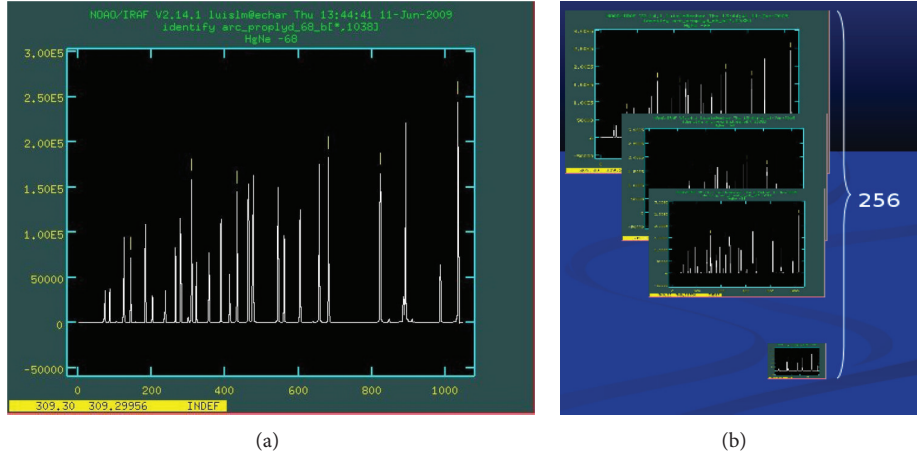

(a)　　　(b)

Figure 5: Wavelength calibration using Cu-Ne-Ar calibration lamp spectra ((a) using *identify*), and application of this wavelength calibration solution to the rest of the spectra ((b) using *reidentify*).

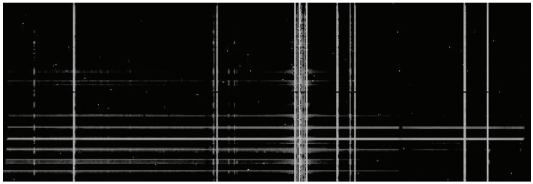

Figure 6: Example of wavelength calibrated spectra image of INTEGRAL data.

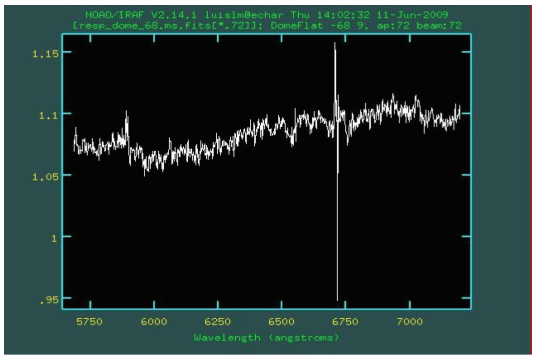

Figure 7: Example of instrumental response for one of the spectra of INTEGRAL data.

This capability is an important improvement compared to the classical long-slit spectroscopy that allowed only performing one-dimensional analysis. We mention here two examples showing that bidimensional spectroscopy is also possible with a long-slit spectrograph, by using a complicated observing setup consisting of taking several long-slit spectra at adjacent positions, shifting the pointing of the telescope by small distances (of the order of the required spatial resolution) with the position angle of the slit perpendicular to the spatial shift of the telescope [9, 10]. The results of this technique were found to be satisfactory although they were not comparable to those of real 3D spectroscopy and the main differences were found in the morphological details of the bidimensional reconstructed images.

In this section we apply the results of the data reduction and calibration process to a particular scientific case: the study of the Herbig-Haro object HH202 in the Orion nebula (see Figure 9). A detailed study on this object and the complete results of this work can be found in Mesa-Delgado et al. [2].

Herbig-Haro objects are nebulosities associated to high-velocity clouds of ionized gas. Several of them have been detected in the Orion nebula (e.g., HH202, HH203, and HH204, among others). The origin of these objects is still uncertain but some authors associate them with infrared objects [11].

In particular, HH202 was observed at Calar Alto Observatory with the fiber-fed spectrograph PMAS [5] at the 3.5 m telescope using lens array configuration, giving a field of view of $16'' \times 16''$, with a spectral resolution of 3.6 Å and a spatial sampling of $1''$ (see Figure 9). The total spectral coverage ranged from 3500 Å in the blue end to 7200 Å in the red end, in two different setups overlapping in the range 5100 Å–5700 Å. This choice of the spectral range ensures the observation of the emission lines of interest.

In this work, two sets of emission lines were selected with different purposes.

Balmer lines (H$\alpha$, H$\beta$,..., HII) were used to estimate the correction for dust extinction (following the theoretical line ratios from [12], and using the reddening function by [13]). These lines were used in addition to correct for differential atmospheric refraction (DAR [14]).

Atmospheric refraction is the deviation of the light as it passes through the atmosphere due to the variation in air density as a function of altitude. Bluer wavelengths are refracted more strongly than redder wavelengths, which means that a white point source is spread out into a little spectrum along the elevation angle producing the called differential atmospheric refraction. In the absence of DAR the images of a star at all the wavelengths are coincident, and the spectra obtained from any aperture are basically the same, but when DAR is present the images of a star at different wavelengths are not in positional agreement. One of the most interesting aspects of IFS is that it is possible to determine and to correct in the spectra the effects of this DAR using a posteriori procedure (e.g., [15]). For this, we use the lines of the Balmer series. The Balmer series is a set of recombination lines corresponding to H$^+$ and spanning a wide range of wavelength (although always in the optical). In the absence of dust, the relative intensities of these lines are fixed, but this is never observed in real nebula. In addition, all these lines should be spatially coincident. But, as mentioned above, the effect of the atmosphere shifts the emission of the different Balmer lines depending on their wavelength. This disadvantage can be sorted out since IFS provides bidimensional maps of any spectral feature of interest. Thus, a simple spatial matching of the observed Balmer emission line maps solves this problem.

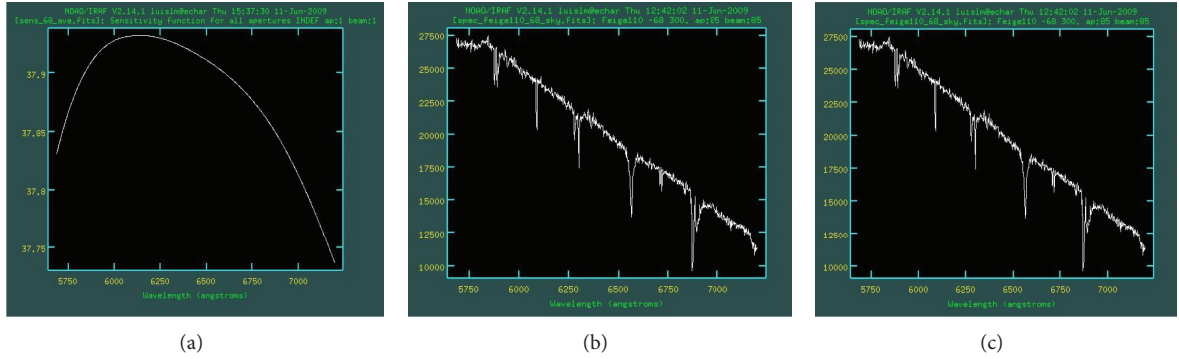

Figure 8: (a) Response function of INTEGRAL data to the different wavelength. (b) Nonflux calibrated spectrum (counts units). (c) Flux calibrated spectrum (flux units).

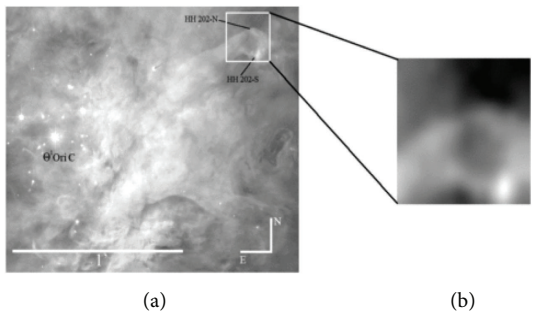

Figure 9: *HST* image of the central part of the Orion Nebula, which combines Wide-Field Planetary Camera 2 (WFPC2) images taken in different filters [4]. The white square corresponds to the FOV of PMAS IFU used, covering the head of HH 202. The separate close-up image on the right shows the H$\alpha$ map obtained with PMAS. The original map is 16 × 16 pixel of 1 × 1 arcsec$^2$ size and has been rebinned to 160 × 160 pixel. Note the remarkable similarity between the *HST* image and our rebinned PMAS H$\alpha$ map. Extracted from Mesa-Delgado et al. [2].

The first step for the correction of the DAR effects is to determine the shifts among images generated at different wavelengths. In our case, we have noticed the effect of the differential atmospheric refraction in the monochromatic images of HH 202 obtained for Balmer lines at different wavelengths reaching the value of ~1.3 arcsec between H$\alpha$ and H11. We have measured offsets between all Balmer line images and shifted with respect to H$\alpha$.

A second set of emission lines was used to estimate abundances and can be split into two categories: (a) collisionally excited lines ([OIII]4363, 4959, 5007 Å, [SII]6717, 6731 Å, [NII]5755, 6548, 6583 Å) to estimate physical properties, like electronic density and temperature and ionic abundances; and (b) recombination lines of CII and OII (and others) to estimate ionic abundances.

To illustrate the power of the 3D spectroscopy technique we focus on an open astrophysical problem discussed in Mesa-Delgado et al. [2]. Figure 10 shows the spatial distribution of the abundances of several ions using different methods: the collisionally excited lines (CELs) and the recombination lines (RLs). Figure 10 stresses the relevance of the abundance discrepancy factor (ADF) which is defined as the difference between abundances determined from CELs and RLs. In particular, Figures 10(a) and 10(b) show clear differences of the same quantity (12 + log O$^{2+}$/H$^+$) depending on the method used to compute it, which in some cases are larger than 0.2 dex. Interestingly, these differences are not constant but vary from point to point, suggesting a complexity that probably involves several physical variables. The ADF problem has been largely studied in the literature for HII regions and for planetary nebulae [16–20] with no clear concluding results about its origin.

The most relevant aspect that 3D spectroscopy can provide to this fundamental astrophysical problem is the fact that all the physical information relevant to tackle on this problem is available for each spatial element: electronic density and temperature, reddening, underlying stellar population, and so forth. This means that a complete analysis can be performed, which will likely result in a particular solution for each spaxel. But even more, the whole set of solutions must keep an internal coherence that must be consistent with the physical assumptions adopted. Moreover, the bidimensional results obtained from 3D spectroscopy must also put constraints to the physically motivated theoretical models of radiation transfer in ionized media.

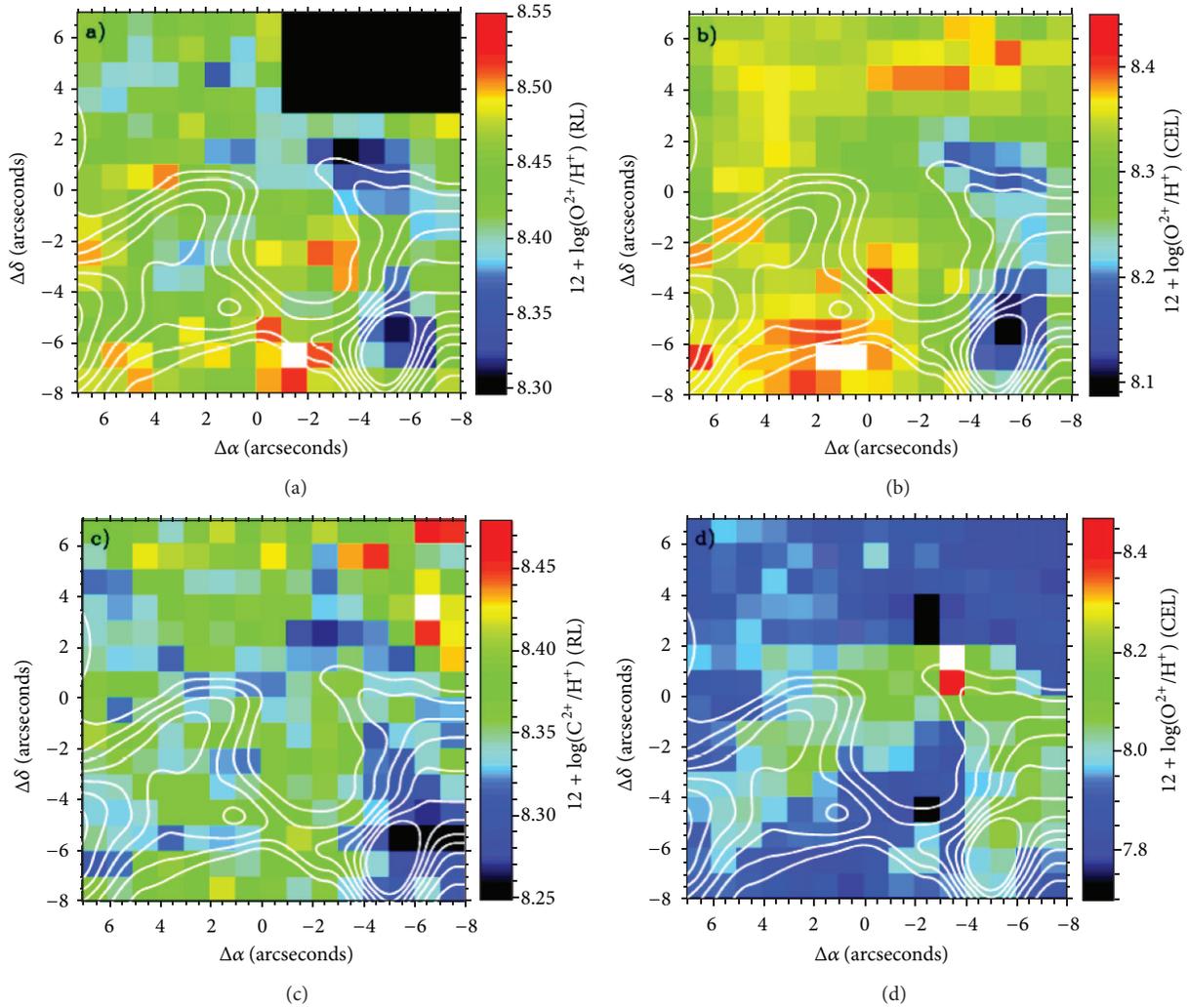

Figure 10: Ionic abundance maps with H$\alpha$ contours overplotted: (a) $12 + \log O^{2+}/H^+$ from RLs, (b) $12 + \log O^{2+}/H^+$ from CELs, (c) $12 + \log C^{2+}/H^+$ from RLs, and (d) $12 + \log O^+/H^+$ from CELs. Extracted from Mesa-Delgado et al. [2].

## Conflict of Interests

The author declares that there is no conflict of interests regarding the publication of this paper.